\documentclass[letter]{jpsj2}

\def\runtitle{Antiferro-quadrupole Ordering of CeB$_6$ Studied 
       by Resonant X-ray Scattering}
\def\runauthor{Hironori {\sc NAKAO}}

\title {Antiferro-quadrupole Ordering of CeB$_6$ Studied 
       by Resonant X-ray Scattering}
\author{Hironori {\sc Nakao},
 \footnote{E-mail: hironori.nakao@kek.jp}
 Ko-ichi {\sc Magishi},$^1$ Yusuke {\sc Wakabayashi},$^{2}$ 
 Youichi {\sc Murakami}, Kuniyuki {\sc Koyama},$^1$ 
 Kazuma {\sc Hirota},$^3$ Yasuo {\sc Endoh},$^4$ and Satoru {\sc Kunii}$^3$}

\inst{
 Photon Factory, Institute of Materials Structure Science, High Energy 
 Accelerator Research Organization (KEK), Tsukuba 305-0801, Japan\\
 $^1$ Faculty of Integrated Arts and Sciences, The University of Tokushima\\
 $^2$ Department of Physics, Keio University, Yokohama 223-8522, Japan\\
 $^3$ Department of Physics, Tohoku University, Sendai 980-8578, Japan\\
 $^4$ Institute for Materials Research, Tohoku University, 
  Sendai 980-8577, Japan\\ }
\recdate{\today}

\abst
{
Under zero magnetic field, a quadrupolar order parameter 
at $\mib{q}_Q=(\frac{1}{2},\frac{1}{2},\frac{1}{2})$
in a typical antiferro-quadrupole (AFQ) ordering compound CeB$_6$
has been observed for the first time 
by means of a resonant X-ray scattering (RXS) technique.
The RXS is observed at the $2p\rightarrow 5d$ 
dipole transition energy of the Ce $L_3$-edge.
Using this RXS technique to observe the pure order parameter of the AFQ state, 
the magnetic phase diagram of Phase II is first determined.
}
\kword
{ CeB$_6$, antiferro-quadrupole ordering, 
 orbital ordering, resonant X-ray scattering, magnetic phase diagram}

\begin{document}

\sloppy
\maketitle

 For both $d$- and $f$-electron systems, various phase transitions 
 involving the orbital degree of freedom are attracting much interest.
 The orbital degree of freedom is usually coupled with atomic displacements
 by the Jahn-Teller effect.
 For example, $3d$ electrons are spread in space so that the orbital 
 ordering is definitely accompanied with lattice distortion.
 In the $f$-electron system, however, 
 orbital ordering  without atomic displacement 
 owing to electron-electron interaction exists,
 which is called antiferro-quadrupole (AFQ) ordering.

\begin{figure}[bht]
 \noindent
 \includegraphics[width=7.5cm,angle=0]{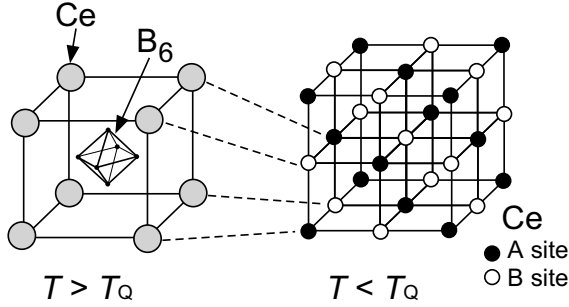}
 \caption[Structure]{
 (left) Crystal structure of CeB$_6$ at $T>T_Q$.
 (right) AFQ structure at $T<T_Q$.
 This is an NaCl-type ordering of Ce A, B sites;
 the structure expects the order parameter 
 at $\mib{q}_Q=(\frac{1}{2},\frac{1}{2},\frac{1}{2})$.
 B$_6$ octahedron is omitted.
 }
 ~\label{fig1}
\end{figure}

 CeB$_6$ has been considered a typical compound for AFQ ordering.%
 ~\cite{ceb_st,gamma8,specific_heat,neutron1,%
  nmr,theory_octa,ultrasonic,fluctuation-th,sample,neutron2}
 CeB$_6$ has a cubic CsCl-type structure with Ce and B$_6$ octahedra
 as shown in Fig.~\ref{fig1}.
 The space group is $Pm\bar{3}m$ and
 the lattice constant is $a=4.14$~\AA .~\cite{ceb_st}
 The ground state of the Ce$^{3+}$ multiplet is a $\Gamma_8$ quartet 
 of the spin doublet (Kramers doublet) and the orbital doublet.~\cite{gamma8}
 The specific heat measurement indicates the anomaly of a sharp peak 
 at $T_N\sim2.3$~K and a broad satellite peak at $T_Q\sim3.2$~K.%
 ~\cite{specific_heat}
 When a magnetic field is applied, 
 $T_Q$ shifts to a higher temperature with a rise in the peak height
 and $T_N$ decreases rapidly.
 This unusual behavior of $T_Q$ has attracted much attention.
 The magnetic scattering of neutron experiments proved
 that the phase below $T_N$ (Phase III) is an antiferro-magnetic state.
 In the phase at $T_N<T<T_Q$ (Phase II), on the other hand, 
 there is no magnetic ordering
 although $T_Q$ is strongly affected by the magnetic field.%
 ~\cite{neutron1}
 Because the entropy calculated from the specific heat
 reaches $R\ln2$ at $T_N$,~\cite{specific_heat} 
 the Kramers degeneracy was expected to lift below $T_N$.
 Based on these experimental results,
 the AFQ transition is expected at $T_Q$,
 although the AFQ order parameter has never been observed
 under zero magnetic field.
 This is a famous hidden orbital ordering of the $f$-electron.

 A magnetic field was applied
 to confirm the orbital ordering in Phase II.
 The field-induced moment was observed 
 by neutron scattering~\cite{neutron1} and NMR.~\cite{nmr}
 The neutron experiment observed the order parameter
 at $\mib{q}_Q=(\frac{1}{2},\frac{1}{2},\frac{1}{2})$
 indicating a NaCl-type AFQ ordering as shown in Fig.~\ref{fig1},
 while the results of the NMR measurements indicated a complicated 
 triple-$\mib{q}$ structure.
 Recently, however, theories including the influence of octapoles
 have clarified that the antiferromagnetic 
 structure of $\mib{q}_Q=(\frac{1}{2},\frac{1}{2},\frac{1}{2})$
 can account for the NMR data.~\cite{theory_octa}
 In ultrasonic measurements, the softening of elastic constants was observed
 toward $T_Q$,
 which implied the AFQ ordering of $\Gamma_5$ symmetry.~\cite{ultrasonic}
 The mysterious behavior of Phase II under the magnetic field was
 theoretically explained by
 the effect of the octupole-octupole interaction~\cite{theory_octa}
 and the fluctuation of the quadrupole moment.~\cite{fluctuation-th}
 All the results suggest the existence of the O$_{xy}$-type AFQ 
 orbital ordering in Phase II, but the AFQ order parameter 
 $\mib{q}_Q$ under zero magnetic field has not been observed up to now.

 Recently, the resonant X-ray scattering (RXS) technique
 to observe orbital ordered states
 has been developed for the  $3d$-electron system.~\cite{murakami1,murakami2}
 When the orbital of $3d$-electrons is ordered,
 the energy level of the $4p$-orbital splits
 owing to some interactions.
 This $4p$-level splitting is the origin of the RXS.~\cite{murakami2}
 Two mechanisms for the $4p$-level splitting are 
 theoretically proposed, however,
 the mechanisms are a controversial problem.~\cite{RXS}
 In this $3d$-electron system, the orbital ordering can also be discussed 
 on the basis of the state of the lattice distortion
 studied by structural analysis.~\cite{kaji}
 In the $4f$-electron system, on the other hand, 
 the RXS is the only technique for measuring the order parameter
 of AFQ without magnetic ordering under zero magnetic field.
 For the $L_3$-edge of the Ce ion, the energy level splitting of
 the $5d$-orbital is expected to induce the RXS.
 The $5d$-level splitting is caused by 
 the Coulomb interaction between $4f$- and $5d$-orbitals
 rather than by the lattice distortion,
 because the localized $4f$-electron has the weaker coupling with 
 the lattice.
 Actually, no superlattice peak due to lattice distortion~\cite{kasuya} 
 has been observed below $T_Q$ in CeB$_6$ up to now.
 In DyB$_2$C$_2$ of the $4f$-electron system, Hirota {\it et al.} 
 successfully measured an RXS signal of the AFQ ordering
 for the first time.~\cite{dybc1}
 In order to develop the RXS technique to 
 observe the orbital ordering in $f$-electron systems,
 a study on AFQ ordering in CeB$_6$ is very important.
 In this letter, we present the AFQ order parameter in CeB$_6$ 
 which has been observed for the first time by using the RXS technique
 under zero magnetic field and a magnetic field up to 2~T.

 A single crystal of CeB$_6$ was fabricated by the 
 floating zone method under pressurized high-purity Ar gas.~\cite{sample}
 The $(1\ 1\ 1)$ surface, $\sim$3$\times$1.5 mm$^2$, was cut
 and polished with fine emery paper.
 X-ray scattering experiments were carried out at the beam lines (BL)
 4C and 16A2 of the Photon Factory in KEK.
 At BL-16A2, the incident beam was monochromatized 
 by Si(111) double crystals;
 the second crystal was used for horizontal sagittal focusing.
 The beam was vertically focused using a mirror.
 The incident X-ray energy 
 is about 5.72~keV of a Ce $L_{3}$-edge 
 and the energy resolution is about 1~eV.
 A four-circle diffractometer equipped with a He-flow cryostat was used.
 The flow of He gas in the sample space prevented the sample surface
 from heating up due to X-ray irradiation.
 The temperature can be decreased to 2.5~K by He pumping,
 and the magnetic field can be applied up to $H=2$~T 
 using a superconducting magnet.
 The polarization vector $\sigma$ of the incident X-ray 
 was parallel to $[11\bar{2}]$ of the crystal structure and
 the magnetic field was also applied along the $[11\bar{2}]$
 in this experiment.

\begin{figure}[tbh]
 \noindent
 \includegraphics[width=7.5cm,angle=0]{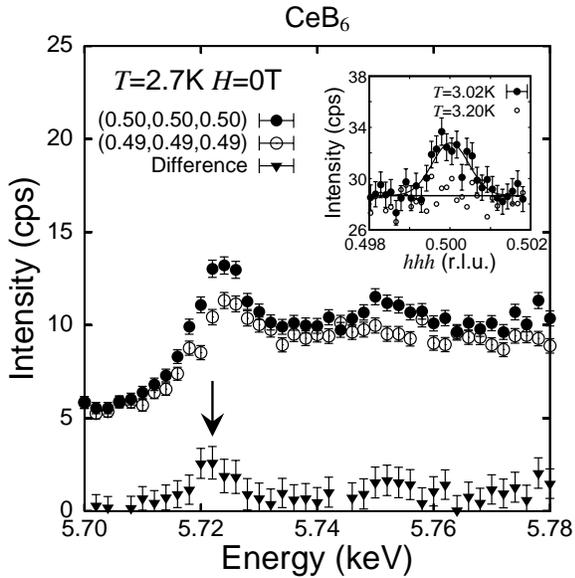}
 \caption[RXS-Edep]{
  Energy dependence of scattering intensities
  at $(\frac{1}{2},\frac{1}{2},\frac{1}{2})$ (filled circles)
  and at $(0.49,0.49,0.49)$ (open circles) with $H=0$~T and $T=2.7$~K.
  The difference of intensities between 
  $(\frac{1}{2},\frac{1}{2},\frac{1}{2})$ and $(0.49,0.49,0.49)$ 
  (filled triangles).\\
  Inset: Temperature dependence of $(hhh)$ scans
  at 5.722~keV and at $H=0$~T.
 }
 ~\label{fig2}
\end{figure}

 The RXS of the order parameter 
 at $\mib{q}_Q=(\frac{1}{2},\frac{1}{2},\frac{1}{2})$
 due to AFQ ordering has been searched under zero magnetic field.
 The energy dependence of scattering intensities
 at $(\frac{1}{2},\frac{1}{2},\frac{1}{2})$
 is measured when $H=0$~T and $T=2.7$~K, as shown in Fig.~2.
 The intensity at $(0.49,0.49,0.49)$ 
 is mainly attributed to the fluorescence of the Ce ion so that
 we regard the intensity as the background.
 The difference between the two intensities 
 is also shown in Fig. 2.
 The weak signal of RXS is observed 
 at the $2p\rightarrow 5d$ dipole transition energy of 5.722~keV
 which is denoted by an arrow.
 The $(hhh)$ scans of the reciprocal lattice space at 5.722~keV
 are shown in the inset of Fig.~\ref{fig2}.
 The half-width at half-maximum of the peak at $T=3.02$~K ($<T_Q$) 
 is 0.00053\AA$^{-1}$ which is 
 resolution-limited in this experimental configuration;
 namely, this phase is a long-range ordered state.
 The dependence of the intensity on the energy and ${\mib q}$-space
 is strong evidence that this peak intensity is attributable to 
 the AFQ ordering in CeB$_6$.
 The signal of RXS disappears at $T=3.20$~K ($>T_Q$),
 as shown in the inset.
 Consequently, the temperature dependence of the AFQ order parameter
 is clearly observed.

\begin{figure}[tbh]
 \noindent
 \includegraphics[width=7.5cm,angle=0]{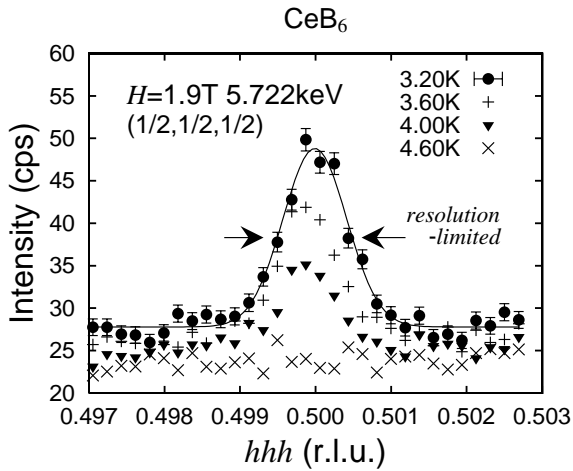}
 \caption[1.9T-th2th-tdep]{
  Temperature dependence of $(hhh)$ scans
  at $E=5.722$~keV and $H=1.9$~T.
 }
 ~\label{fig3}
\end{figure}

\begin{figure}[t]
 \noindent
 \includegraphics[width=7.5cm,angle=0]{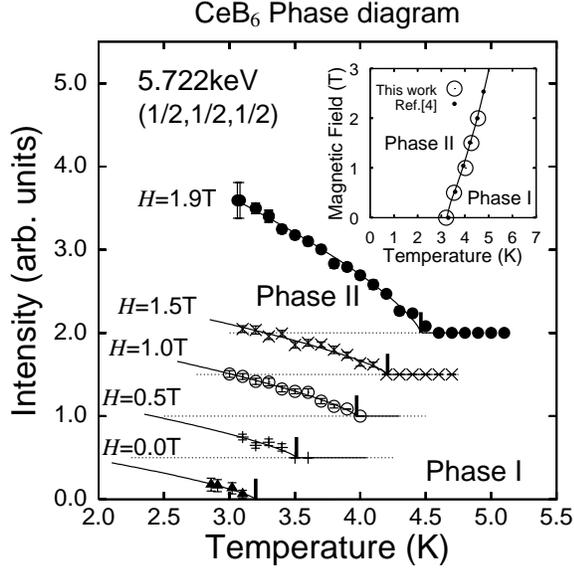}
 \caption[Phase diagram]{
 Temperature dependence of RXS intensities of CeB$_6$ 
 with various magnetic fields, $H\parallel [11\bar{2}]$.
 The solid lines are results of fitting with the critical index $\beta=0.37$.
 The transition temperatures $T_Q$ are indicated by thick vertical lines.
 The baselines are shifted for clarity\\
 Inset:   $H-T$ phase diagram for the present result
 and the result of the neutron experiment
 with the magnetic field $H\parallel [111]$ (Ref.[4]).
 }
 ~\label{fig4}
\end{figure}

 Next, we investigate the AFQ ordering in a magnetic field
 using this RXS technique.
 Figure \ref{fig3} shows the temperature dependence 
 of the RXS of $(\frac{1}{2},\frac{1}{2},\frac{1}{2})$
 at $E=5.722$~keV and $H=1.9$~T.
 The RXS intensity is much larger than that of the RXS 
 under zero magnetic field.
 The peak profiles can be fitted by Gaussian curves
 and the peak widths are also resolution-limited.
 The dependence of the RXS integrated intensities
 on the temperature and field is summarized in Fig.~\ref{fig4}.
 As the magnetic field is increased, the RXS intensity
 becomes larger and the transition temperature $T_Q$ shifts to higher ones.
 The intensities of RXS  at low temperatures grow
 with increasing field owing to the development of the order parameter 
 of the AFQ state.
 These temperature dependences of the intensities are fitted well by
 $I_{(\frac{1}{2},\frac{1}{2},\frac{1}{2})} \propto
 \bigl(\frac{T_Q-T}{T_Q}\bigr)^{2\beta}$
 with the critical exponent $\beta=0.37$
 which was determined by the neutron scattering experiment.~\cite{neutron2}
 The solid lines in Fig.~\ref{fig4} indicate this function.
 All the data can be fitted by this function with the same $\beta$ 
 within experimental error.
 Based on this fitting, the $H-T$ phase diagram is determined,
 as shown in the inset of Fig.~\ref{fig4}.
 This phase diagram is in good agreement 
 with previous result of neutron diffraction.~\cite{neutron1}
 The increase in $T_Q$, which is an unusual behavior of CeB$_6$, is clearly 
 shown in the phase diagram.
 All the peak widths below $T_Q$ are resolution limited
 and have no temperature dependence.
 The short-range ordered state could not be observed 
 at $T>T_Q$ owing to a weak signal although
 the short-range ordering is expected theoretically.~\cite{fluctuation-th}
 Thus, by using the RXS technique at the Ce $L_3$-edge
 $2p\rightarrow 5d$ dipole transition energy,
 the order parameter of AFQ has been measured
 and the phase diagram has also been determined.
 It is noted that the RXS technique is the only experimental technique
 for observing the order parameter of AFQ  under zero magnetic field.

 When a magnetic field is applied, X-ray scattering,
 as well as neutron scattering, can be used to detect
 magnetic scattering of the field-induced moment.
 Therefore we should pay attention to the origin of the observed intensity.
 Under zero magnetic field, on the other hand, 
 the intensity of RXS corresponds not to magnetic scattering,
 but to the pure AFQ order parameter.
 The scattering intensity does not change much when the field is applied.
 Therefore, it is expected that the RXS under magnetic field also
 corresponds to the pure order parameter of AFQ ordering and 
 has no component of magnetic scattering.

 On the basis of the azimuthal angle dependence of RXS,
 the wave function of the ordered orbital can be determined quantitatively,
 which has been demonstrated in the case of YTiO$_3$.~\cite{ytio_nakao}
 We attempted to measure the azimuthal angle dependence of RXS
 under zero magnetic field.
 The intensity was so weak in CeB$_6$ that
 a quantitative discussion could not be made.
 However, it is possible to determine 
 the wave function of the ordered orbital in the AFQ state
 based on the azimuthal angle dependence of RXS.
 Moreover, there is a possibility of determining
 the various types of the induced quadrupole ordered state 
 depending on the applied field directions,
 which have been predicted by some theories.~\cite{theory_octa}
 This RXS at the dipole transition energy is represented by a tensor of 
 the atomic scattering factor, which only reveals the symmetry 
 of the quadrupole ordered state.
 On the basis of RXS at the dipole transition, therefore, 
 we unfortunately can not observe octapole.

 The RXS observed here is considered to reflect the energy level split 
 of the Ce $5d$-orbital due to
 the Coulomb interaction between $4f$- and $5d$-orbitals,
 however, the scattering mechanism is still a controversial problem.
 In fact, we also must consider an anisotropic potential
 of the core $2p$-hole.
 Development of a theory concerned with the RXS mechanism 
 in the $f$-electron system is strongly desired.

 We have succeeded in measuring the order parameter
 $\mib{q}_Q=(\frac{1}{2},\frac{1}{2},\frac{1}{2})$ 
 under zero magnetic field
 in a typical AFQ-ordered system, CeB$_6$, for the first time.
 The RXS is observed at the $2p\rightarrow 5d$ dipole transition energy
 of the Ce $L_3$-edge owing to
 the energy level split of the Ce $5d$-orbital.
 Only the RXS technique can be used to observe the order parameter
 including information on Q of AFQ ordering under zero magnetic field,
 as opposed to neutron scattering observing the field-induced moment.
 Using this observation of the pure order parameter of AFQ,
 the magnetic phase diagram is elucidated.
 This success with the typical AFQ compound, CeB$_6$,
 will be crucial in the study of AFQ materials.

The authors would like to thank Professor Hishizume for the use
of the He-flow-type cryostat with a superconducting magnet.
This work was supported by Core Research for Evolutional Science
and Technology Corporation (CREST).
This study has been performed under the approval of the Photon
Factory Program Advisory Committee (Proposal No. 99G211).


\end{document}